\documentclass[fleqn,12pt,twoside]{article}
\usepackage[headings]{espcrc1}
\readRCS
$Id: espcrc1.tex,v 1.2 2004/02/24 11:22:11 spepping Exp $
\usepackage{graphicx}
\usepackage[figuresright]{rotating}

\newcommand{\AmS}{{\protect\the\textfont2
  A\kern-.1667em\lower.5ex\hbox{M}\kern-.125emS}}
\title{The background for Cherenkov gluons at RHIC and LHC energies}

\author{I.M. Dremin\address{Lebedev Physical Institute, Moscow (Russia)}
        \thanks{Email: dremin@lpi.ru},
        L.I. Sarycheva\address[SINP]{D.V. Skobeltsyn Institute of Nuclear Physics,\\
                 M.V. Lomonosov Moscow State University, Moscow (Russia)}
	\thanks{Email: lis@alex.sinp.msu.ru},
        K.Yu. Teplov\addressmark[SINP]
	\thanks{Email: teplov@lav01.sinp.msu.ru}
}  

\runauthor{I.M. Dremin et al.}

\begin{document}

\maketitle

\begin{abstract}
The pseudorapidity distribution of centers of dense isolated groups
of particles in\break HIJING model is determined. It can be considered as the
background for Cherenkov gluons. If peaks over this background were found
in experiment, they would indicate the onset of new collective effects.
\end{abstract}

\bigskip

The search for collective effects in hadronic and nuclear high-energy
reactions has always been one of the mainstreams of experimental and 
theoretical investigations. Among them, Cherenkov gluons \cite{d1,d2} 
and Mach waves \cite{glas} were discussed a long time ago. Recently, the
interest to them was revived \cite{dim,mwa,kmwa,shur,stoc,mrup} in 
connection with RHIC data \cite{wang} as reviewed
in \cite{dim}. The common feature of these collective coherent processes
is particle production concentrated on a cone with the polar angle 
$\theta $ defined by the condition
\begin{equation}
\cos \theta = \frac{c_w}{v}~,     \label{cos}
\end{equation}
if the infinite medium at rest is considered. Here $v$ is the velocity of the 
particle producing these waves, $c_w$ is the phase velocity of gluons
or sound in the medium. At high energies $v\approx c$ and $c_w$ ranges 
from quite low values to a value slightly below $c$ according to present 
estimates for both effects (see \cite{dim}). The lowest values of $c_w$ are 
obtained for dilute media and low energy gluons, while larger $c_w$ 
correspond to strong shock waves and high energy gluons. For gluons 
$c_w=c/n$ where $n$ is their nuclear index of refraction. This index was
estimated from experimental data on hadronic reactions \cite{d2,dim}.
The finite length of nuclear targets can change the estimate (\ref{cos})
\cite{d2,dim}.

The main experimental signature of both effects would be two peaks in 
the pseudorapidity distribution of particles produced in high energy 
nuclear collisions which are positioned in accordance with Eq. (\ref{cos}).
The most picturesque image of these effects is the ring-like structure
of events in the plane perpendicular to the direction of propagation
of initiating them body. 

The recently observed at RHIC \cite{wang} effect with two peaks in angular 
distribution about the direction of propagation of the companion jet was
interpreted in \cite{shur} as Mach waves with $c_w=0.33c$. In terms of
Cherenkov gluons, it could be the emission of low energy gluons with nuclear 
index of refraction equal to $3$.

Beside low energy gluons, the forward moving very high energy partons can 
produce high energy Cherenkov gluons. They would result in two peaks of the
pseudorapidity distribution positioned at large c.m.s. angles. The emission
angle in the target rest system is small but much larger than bremsstrahlung 
angles. There are numerous experimental indications in 
favor of this effect (see review in \cite{dim}). The first one of them
was presented in \cite{addk}.

The important problem of experimental search for this effect is the
background due to ordinary processes. Its influence should be minimized.
For doing this we propose to use the distinctive feature of production of
high energy Cherenkov gluons. Namely, each gluon should produce a jet of 
particles which can be distinguished as a high density isolated group
of particles. Separating such groups in experimental data one would 
increase the share of jets produced by Cherenkov gluons among all 
particles. By this choice we omit weakly correlated particles. 
Statistical fluctuations and hard QCD-jets are still accounted but their
probability is lowered and pseudorapidity distribution is rather smooth.
Therefore the role of background in the distribution of the centers of 
such groups becomes lower compared to the overall pseudorapidity 
distribution. Peaks corresponding to Cherenkov gluons should be more 
pronounced. If the peaks in the pseudorapidity plot of the centers of 
separated groups are found in experiment and fit the condition 
(\ref{cos}), then it favors the hypothesis about Cherenkov gluons. 
The positions of the peaks reveal the properties of hadronic matter.

To estimate the background we have used the HIJING model for central
collisions ($b=0$) for Au+Au collisions at RHIC energy
$\sqrt s$=200A GeV and for Pb+Pb collisions at LHC energy $\sqrt s$=5500A GeV.
3500 events were generated in each case.

\begin{figure}[htb]
\begin{minipage}[t]{80mm}
%\framebox[79mm]{\rule[-26mm]{0mm}{52mm}}
\includegraphics[width=18pc]{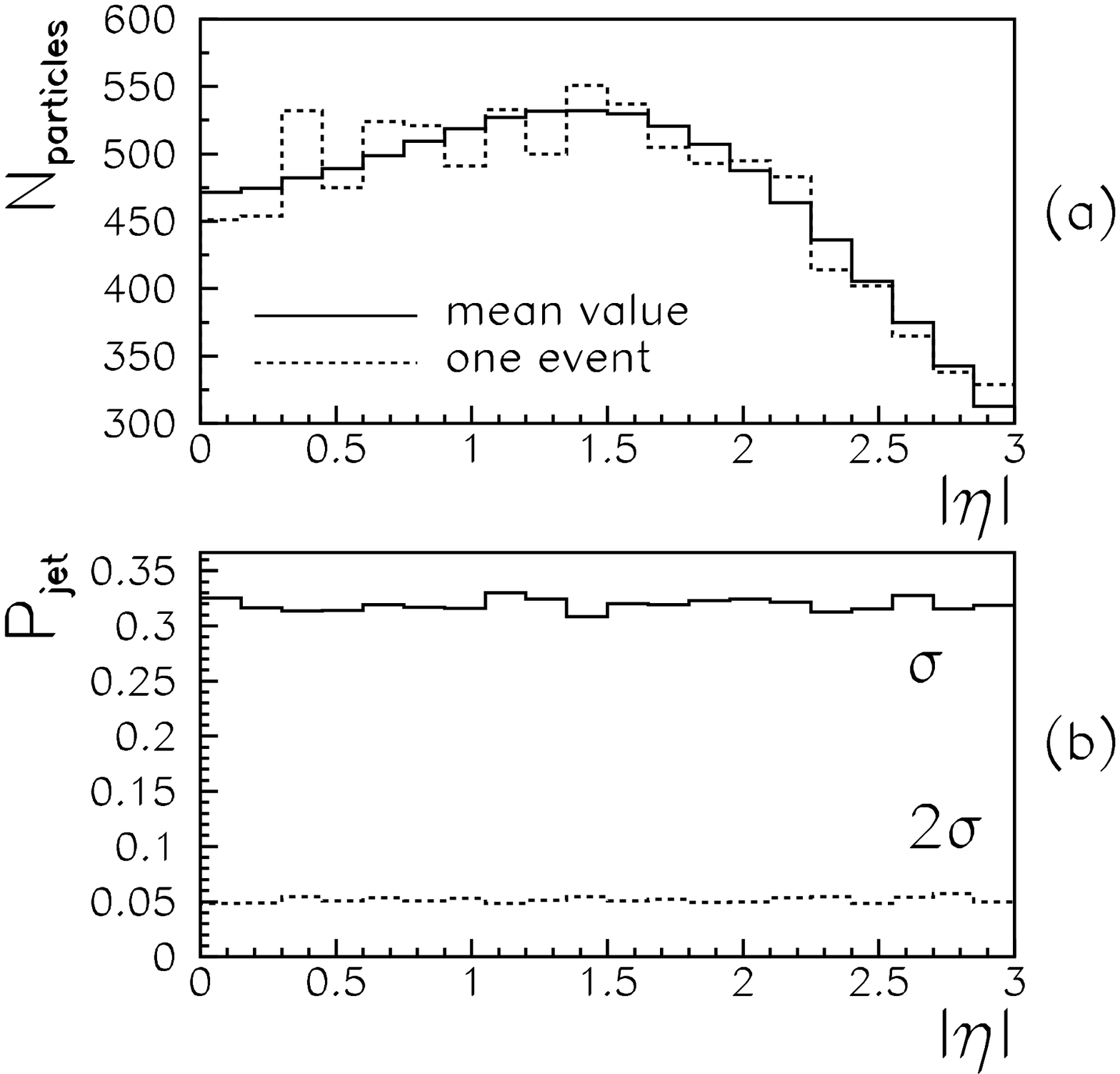}
\caption{(a) The pseudorapidity distribution in one of HIJING events 
(dashed histogram) for central Au+Au collision at $\sqrt s$=200A GeV is plotted 
over the inclusive HIJING distribution (solid histogram), 
 $N_{particles}$ --- number of particles. Peaks over the 
inclusive plot are clearly seen.
(b) The pseudorapidity distribution of the centers of dense
isolated groups of particles similar to those shown in Fig. 1a and 
exceeding the inclusive plot by two and one standard 
deviations $\sigma$, $P_{jet}$ --- probability to find peak
over mean + $\sigma$ ($2\sigma$).
This is the smooth background for further searches
of collective effects.}
\label{fig:pic1}
\end{minipage}
\hspace{\fill}
\begin{minipage}[t]{75mm}
%\framebox[74mm]{\rule[-26mm]{0mm}{52mm}}
\includegraphics[width=18pc]{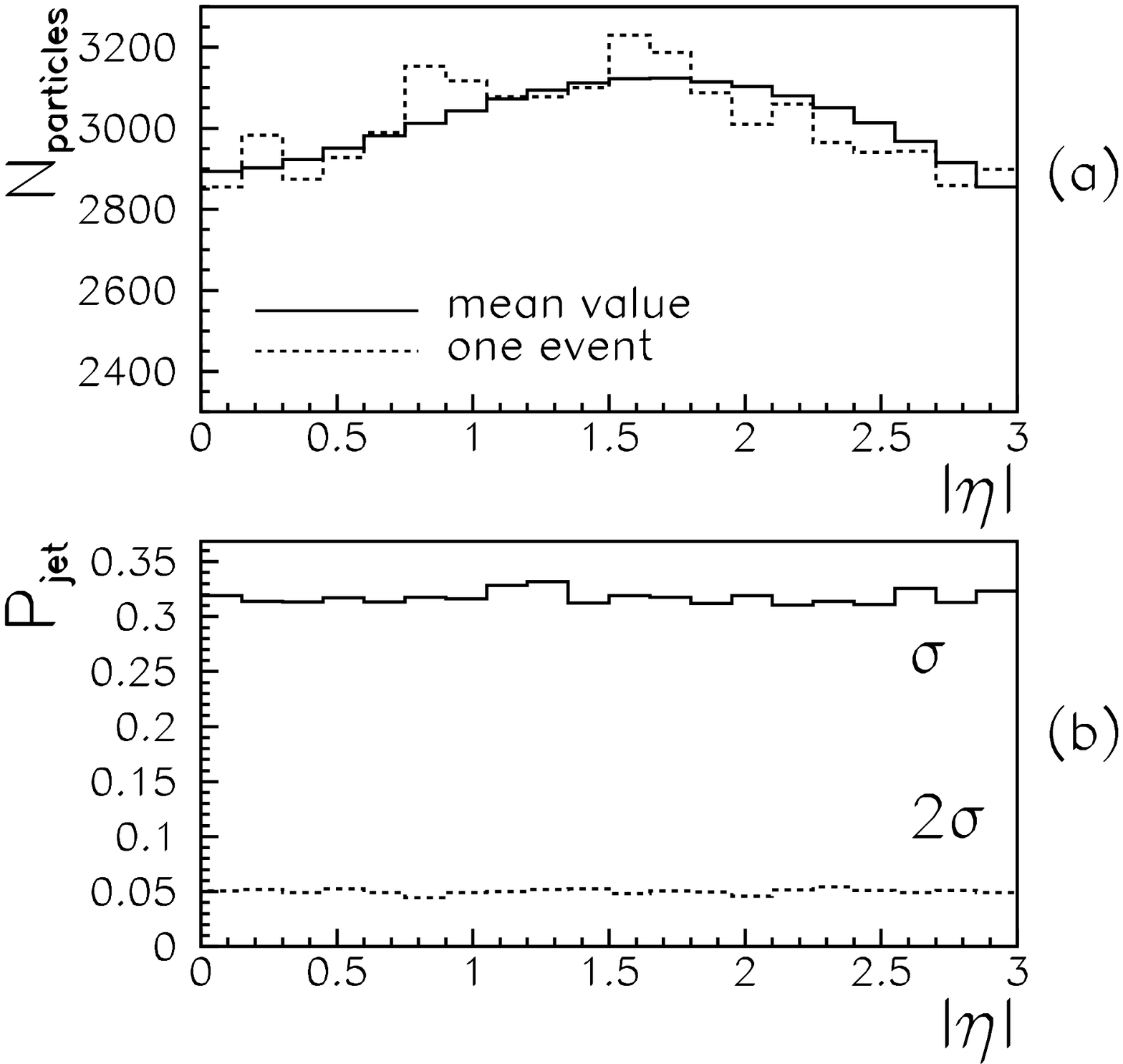}
\caption{(a) The pseudorapidity distribution in one of HIJING events 
(dashed histogram) for central Pb+Pb collision at $\sqrt s$=5500A GeV is plotted 
over the inclusive HIJING distribution (solid histogram), 
 $N_{particles}$ --- number of particles. Peaks over the 
inclusive plot are clearly seen. (b) The pseudorapidity distribution of the centers of dense
isolated groups of particles similar to those shown in Fig. 2a and 
exceeding the inclusive plot by two and one standard 
deviations $\sigma$, $P_{jet}$ --- probability to find peak
over mean + $\sigma$ ($2\sigma$). This is the smooth background for further searches
of collective effects.}
\label{fig:pic2}
\end{minipage}
\end{figure}

Then the spikes in individual HIJING events exceeding this distribution
by more than one and two standard deviations have been separated. They
can appear either as purely statistical fluctuations or as hard QCD-jets.
Figs \ref{fig:pic1}a and \ref{fig:pic2}a show the examples of such events (each one for RHIC and 
LHC energies, correspondingly) plotted over the smooth inclusive
pseudorapidity distributions. 

Peaks exceeding the distributions are clearly seen. All simulated events
have been plotted in such a way and centers of peaks defined.
Finally, the distribution of the centers of these peaks is plotted. 
Figs \ref{fig:pic1}b and \ref{fig:pic2}b show these distributions for peaks exceeding the inclusive plot
at RHIC and LHC energies by two or one standard deviations. It is seen that
these distributions are flat with extremely small irregularities. This
appeals to our expectations that statistical fluctuations and QCD jets do
not have any preferred emission angle. They can be considered  as background 
plots for experimental search for Cherenkov gluons which do have such 
preferred angle. If experimental data on group centers distribution show 
some peaks at definite pseudorapidity values over this background,
this can be an indication on new collective effect, not considered
in HIJING. These findings will add to those experimental facts in favor 
of this effect which existed before (they are reviewed in \cite{dim}).

To conclude, the pseudorapidity distributions of the centers of dense 
isolated groups of particles (jets) exceeding in individual events the 
inclusive distribution are plotted for events generated according to HIJING 
model at RHIC and LHC energies. They provide the background for
further searches for such collective effects as Cherenkov gluons and
Mach waves.\\

This work has been supported in part by the RFBR grants 03-02-16134, 
04-02-16445-a, 04-02-16333, NSH-1936.2003.2.\\

\end{document}